\documentclass{elsart}

\usepackage{graphicx,natbib,amssymb,lineno,hyperref}
\usepackage{lscape}

\newcommand\apj{{ApJ}}%
\newcommand\aj{{AJ}}%

\begin{document}

\begin{frontmatter}
\title{Exoplanet Transit Database. Reduction and processing of the photometric data of exoplanet transits.}

\author[SP,SD]{Stanislav Poddan\'y} \ead{poddany@observatory.cz}
\author[CAS,LB]{Lubo\v{s} Br\'at}
\author[CAS,OP]{Ond\v{r}ej Pejcha}

\address[SP]{Astronomical Institute, Faculty of Mathematics and Physics,
 Charles University Prague, CZ-180 00 Prague 8, V Hole\v{s}ovi\v{c}k\'ach 2, Czech Republic}
\address[SD]{ \v{S}tef\'anik observatory, CZ-118 46 Prague 1, Pet\v{r}\'in 205, Czech Republic}
\address[CAS]{Variable Star and Exoplanet Section of Czech Astronomical Society}
\address[LB]{Altan observatory, CZ-542 21 Pec pod Sn\v{e}\v{z}kou 193, Czech Republic}
\address[OP]{Department of Astronomy, Ohio State University, Columbus, OH 43210, USA}

\begin{abstract}
We demonstrate the newly developed resource for exoplanet
researchers - The Exoplanet Transit Database. This database is
designed to be a web application and it is open for any exoplanet
observer. It came on-line in September 2008. The ETD consists of
three individual sections. One serves for predictions of the
transits, the second one for processing and uploading new data from
the observers. We use a simple analytical model of the transit to
calculate the central time of transit, its duration and the depth of
the transit. These values are then plotted into the observed -
computed diagrams (O-C), that represent the last part of the
application.
\end{abstract}

\begin{keyword}
exoplanets \sep planetary systems \sep techniques: photometric \sep
database

\PACS 95.80.+p \sep 97.82.-j \sep 97.82.Cp

\end{keyword}

\end{frontmatter}

\section{Introduction}

Research on extrasolar planets is currently one of the most exciting
fields in astrophysics. The speculations on the existence of
 planets orbiting other solar-type stars ended fourteen years ago. In
1995 the discovery of the first extrasolar planet orbiting a
solartype star - the well-known 51 Peg b, was made by
\cite{Mayor1995}. Since then, the number of known planets has been
growing quickly. Currently, more than 370 such bodies are
known\footnote{see the list http://www.exoplanet.eu}.

If a planetary system happens to be oriented in the space so that
the orbital plane is close to the line-of-sight to the observer, a
planet periodically passes in front of the stellar disk. Photometric
observation of the transit can then be used to derive the orbital
and physical parameters of the planet (e.g., \citealt{south} or
\citealt{torres}). For a review of properties that have been
measured, or that might be measured in the future through precise
observations of transiting planets see \cite{Winn2008}. At the half
of the year 2009, more than 60 planets with this special orientation
were known.

There are many amateur astronomers all over the world that achieved
photometric accuracy around the units of percent, which is necessary
for quality observing of the exoplanet transit. Unfortunately, up to
present day, no exoplanet light curves' database was available that
would accept data from both professionals as well as from amateurs.
Amateur observers are not constrain by telescope scheduling and
often have unlimited access to their instrument which enables them
to gather data over a long period.

Huge quantity of observations is the key to the search for other
planets in already known systems. It is important to  monitor
possible periodical changes in O-C plots of the planets because as
\cite{Holman2005} demonstrated in their theoretical work, short-term
changes of the time of the transit can be caused by the presence of
other exoplanets or moons in the system, see also \cite{Agol2005},
\cite{Kipping2009}. On the other hand, potential long-term changes
in the duration of the transit may be the consequence of orbital
precession of exoplanets as \cite{Miralda2002} showed in his
theoretical work. To perform such effective studies, we need a
database which includes all available data divided into groups
according their quality.

\section{Why ETD}
\begin{figure}
\includegraphics [width=12cm]{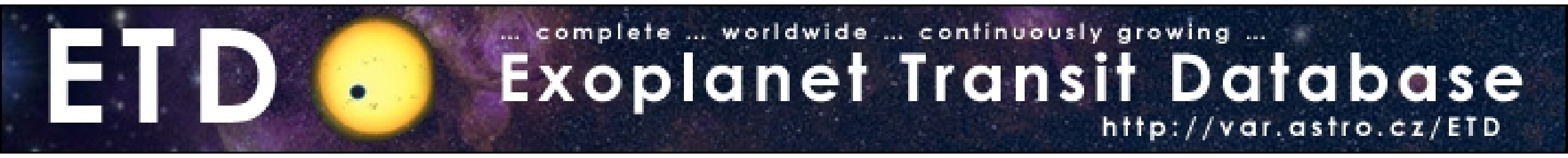}
\end{figure}

The Exoplanet Transit Database\footnote{http://var2.astro.cz/ETD}
(ETD) came on-line in September 2008 as a project of the Variable
Star and Exoplanet Section of the Czech Astronomical Society. The
ETD includes all known transiting planets that have published
ephemerides.

We have created on-line ETD portal to supply observers with such
useful information like transit predictions, transit timing
variation (TTV), variation of depth and duration with availability
to draw user observation to the plot.

Before the ETD, only two other transit databases were available. The
Amateur Exoplanet Archive\footnote{http://brucegary.net/AXA/x.htm}
(AXA), lead by Bruce Gary, and the NASA/IPAC/NExScI Star and
Exoplanet Database\footnote{http://nsted.ipac.caltech.edu} (NStED)
(\citealt{NSTED}). AXA strictly accepts data only from amateurs.
Unfortunately, the quality of light curves is very diverse. In spite
of this fact, all of the available light curves have the same
priority grading. On the other hand NStED contains only the light
curves that have been already published (in the future some amateur
light curves from AXA should be also accepted into the NStED and the
AXA should expire).

The main goal of the ETD is to gather all available light curves
from professional and also amateur astronomers (after one year of
the existence of the database, more than one thousand such records
are available). We are searching for new publications on several
open archives to gather all available light curves. We also take
over data from the NStED, AXA and from our project
TRESCA\footnote{http://var2.astro.cz/EN/tresca}. It is also possible
to upload data into the database directly using a web-form or it can
be added to the database by the administrators. All available data
are on-line plotted into graphs where we make the provision for the
quality of the light curve. All graphs (like TTV) can be easily
downloaded from the database. It is also possible to download light
curves from the TRESCA observers and from amateur observers directly
from the database.

While collecting published data to ETD, we accentuate to have fully
referenced its source. Each transit observation we store full
reference with URL pointing to the paper or web-page where data were
found. When we take over the whole light curve we display it only
with reference to source of the data there is no mention of ETD in
the picture.

\section{Parts of the ETD}
The ETD is composed from three sections. The first one - Transit
prediction - serves for prediction of the transits. The second one -
Model fit your data - is a web-form for accepting and processing new
data. The last section - O-C plots - contains the observed -
computed diagrams of the central times of transits, depths and the
transit durations that are generated on-line from the database.

\subsection{Transit prediction}
This section of the ETD contains one month prediction of observable
transits (the starting day is the date two days before current date)
and also the prediction for the next 365 days for selected
exoplanet. Any observer can find here the time of the transit
start/center/end, duration and the depth of each transit for any
place in the world. Furthermore the altitude and the cardinal point
of the object in the sky are displayed for the first contact,
mid-transit time and the last contact. In the one year prediction
window you can also see the finding chart\footnote{downloaded from
the http://archive.stsci.edu/dss/index.html}
 ($15'$ x $15'$).

\subsection{Model-fit your data}

\begin{figure}
\includegraphics [width=12cm]{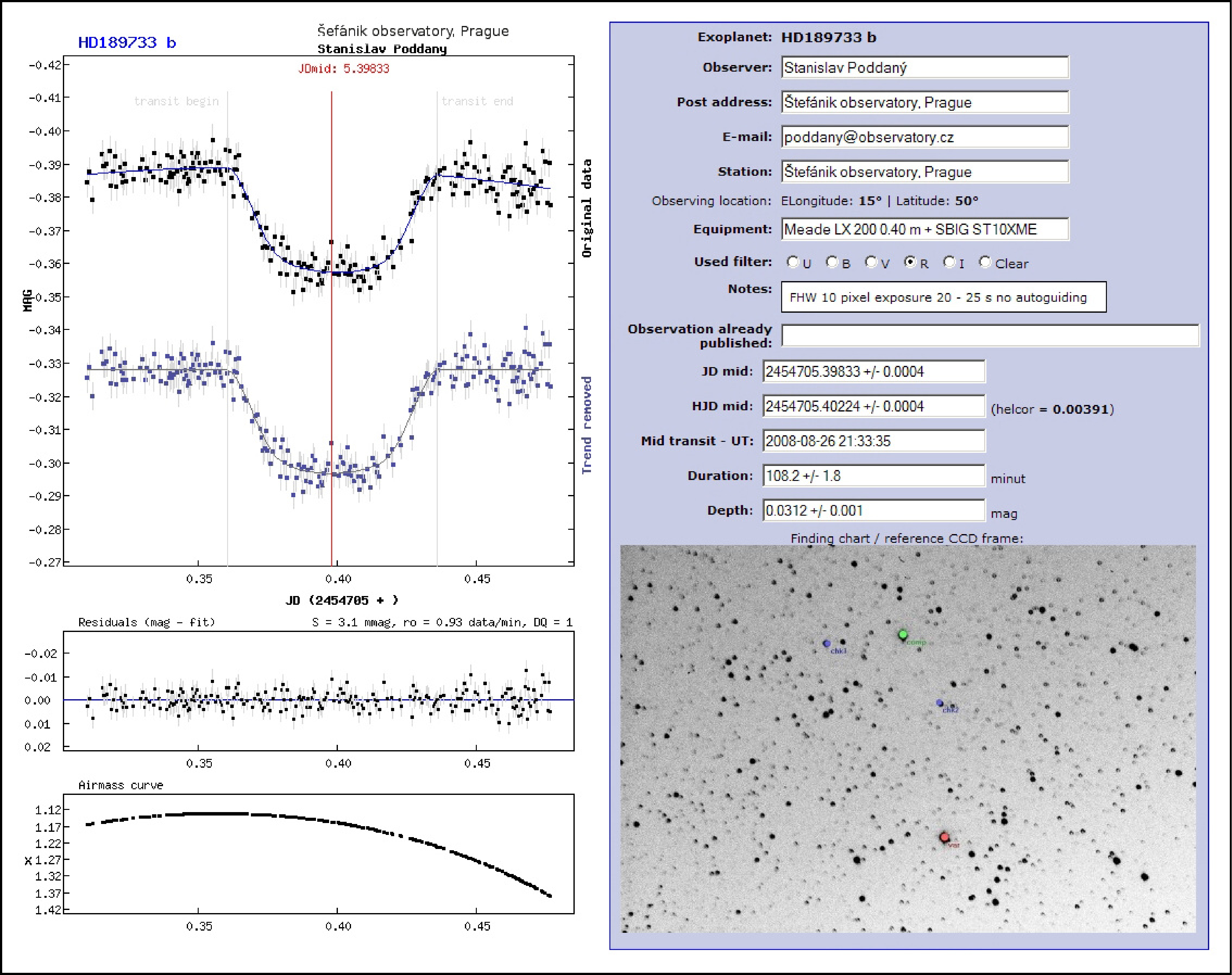} \caption{A record sample in the ETD
after successful processing the light curve.\label{record}}
\end{figure}

This section describes an user-friendly web-form for uploading and
processing the light curves into the database. To model-fit the
transit we assume that the observations consist of $N$ relative
magnitudes $m_{i}$ taken at times $t_{i}$ ($i=1, 2 ... , N$) and the
photometry software provided measurement errors $\sigma_{i}$
computed most likely from Poisson statistics and read-out noise. We
model the dataset by a function
%
\begin{equation}
    \label{rov}
     \begin{array}{r}
    m(t_{i})=A -2.5\log F\left(z\left[t_{i},t_{0},D,b\right],p,c_{1}\right)  \\
    + B(t_{i}-t_{\rm mean})
    + C(t_{i}-t_{\rm mean})^{2},
     \end{array}
\end{equation}
where $F(z, p, c_{1})$ is a relative flux from the star due to the
transiting planet. We assume that the planet and the star are dark
and limb darkened disks, respectively, with radius ratio
$p=R_{P}/R_{\ast}$ and that the planet is much smaller than the
star, $p\lesssim 0.2$. The projected relative separation of the
planet from the star is $z$. Limb darkening of the star is modeled
by the linear law with the coefficient $c_{1}$. We employ the
\textsf{occultsmall} routine of \cite{MA2002} as our $F(z, p,
c_{1})$. We checked that the small planet approximation, $p\lesssim
0.2$, does not produce significant differences from the full model
(at least for the typical values of $p$ and having in the mind the
typical quality of the photometry) and is much faster to compute,
which is the most important factor for on-line processing.

We model the planet trajectory as a straight line over the stellar
disk with impact parameter $ b=a \cos I/R_{\ast}$, where $a$ is a
semi-major axis and $I$ is the orbit inclination. The mid-transit
occurs at $t_{0}$ and the whole transit lasts $D$. Based on these
assumptions we can compute $z[t_{i}, t_{0}, D, b]$ for every
$t_{i}$.

Variable $A$ in the equation (\ref{rov}) descibes the zero-point
shift of the magnitudes, while $B$ and $C$  describe systematic
trends in the data. Linear and quadratic terms are computed with
respect to the mean time of observations $t_{\rm mean}=\sum t_{i}/N$
to suppress numeric errors. We do not employ any explicit correction
for air-mass curvature as we think a generic second-degree
polynomial is sufficient in most cases.

We used the Levenberg-Marquardt non-linear least squares fitting
algorithm from \cite{Press1992}, procedure \textsf{mrqmin}. The
algorithm requires initial values of parameters and partial
derivatives of the fitted function. We take the initial values from
literature (except for $c_{1}$, see below). We compute all partial
derivatives of the equation (\ref{rov}) analytically, except for
$\partial F/
\partial z$, $\partial F/ \partial p$ and $\partial F/ \partial c_{1}$ which were computed numerically using
Ridders' method (procedure \textsf{dfridr} of \citealt{Press1992}).

The search for the optimal parameters is done by iterating the
fitting procedure until the $\Delta\chi^{2}$ (between fits) does not
change significantly. Usually, with good initial values, about ten
iterations are sufficient. Then the error bars $\sigma_{i}$ are
re-scaled to make the final $\chi^{2}=N-g$, where $g$ is a number of
free parameters, and we re-run the fitting procedure to obtain final
errors of the parameters. Original photometric errors are usually
underestimated and this procedure yields more reasonable errors of
the output parameters.

In the optimal circumstances, one would consider all variables in
equation (\ref{rov}), namely $A$, $B$, $C$, $t_{0}$, $D$, $b$, $p$
and $c_{1}$ free parameters. However, these parameters are
correlated to some extent and noisy photometry from a small amateur
telescopes does not permit recovery of all of them. We need to fit
the zero-point shift $A$ and in most cases also a linear systematic
trend $B$. Our primary goal is to get the central time of the
transit $t_{0}$, duration $D$ and depth of the transit $\delta$.
Hence, we set $t_{0}$ and $D$ as free parameters, by default.
However, for a limb-darkened star, the depth of the transit $\delta$
is determined by radius ratio $p$, impact factor $b$ and limb
darkening coefficient $c_{1}$. Primarily, the depth of transit
$\delta$ is governed by the radius ratio $p$ and we set it as a free
parameter. The parameters $b$ and $c_{1}$ affect the depth and the
shape of the transit to a lesser extent and from noisy amateur data
we couldn't retrieve meaningful values for the two parameters
simultaneously with $p$. Therefore, we hold $b$ and $c_{1}$ fixed.
We either compute $b$ from orbital parameters of the planet and from
the radius of the star or take the value from the literature. The
situation is more complicated for limb darkening because $c_{1}$
should be different for every photometric filter. We decided to keep
$c_{1}$ fixed at a rather arbitrary value $c_{1}=0.5$ in all cases.
We experimented with values from $0.2$ to $0.9$ and found that the
effect on other parameters is rather negligible, usually smaller
than the error bars. The export value of the depth is then evaluated
as
\begin{equation}
\delta=-2.5 \log\left[\min\limits_z F\left(z, p,
c_{1}\right)\right].
\end{equation}
At Fig. \ref{record} you can see the example of the record in the
ETD after successful processing of the light curve.

We tested our algorithm using the data of the exoplanet HD189733b
that were published by \cite{Winn2007}. When we fitted this precise
data using our code we obtained the value of the mid-transit time
which was in excellent agreement with the published results. The
duration of the transit wasn't inside the error bars. The errors of
our fit are lower than those in the paper where the MCMC simulations
were used (Tab. \ref{Table1}). We think that this excess is due to
red noise which we did not take into account in our calculation and
because of impact parameter that is fixed in our case. We also made
a test with the data from AXA and we obtained similar results to the
results presented in their archive (Tab. \ref{Table2}).

\begin{table}
\caption[]{The comparison of our results with \cite{Winn2007}.}
\label{Table1}
\centering \scalebox{1}{                        
\begin{tabular}{c c c }
\hline \hline \noalign{\smallskip}
    HD189733b   & ETD & \cite{Winn2007}    \\
    \hline \noalign{\smallskip}
    central time [HJD] &    2453988.80333 (12)     &    24543988.80331 (27)     \\
    \noalign{\smallskip}
   duration [min.] &  106.01 $\pm$ 0.50 & 109.62 $\pm$ 1.74        \\
    \hline
\end{tabular}}
\end{table}

\begin{table}
\caption[]{The comparison of our results with the AXA database.}
\label{Table2}
\centering \scalebox{1}{                        
\begin{tabular}{c c c }
\hline \hline \noalign{\smallskip}
    HD189733b   & ETD & AXA    \\
    \hline \noalign{\smallskip}
    central time [HJD] &    2454705.40228  (41)     &    2454705.4023 (5)     \\
    \noalign{\smallskip}
    duration [min.] &  102.09 $\pm$ 1.60    &    98.4 $\pm$ 1.8      \\
    \noalign{\smallskip}
    depth [mag.] &  0.0287  $\pm$ 0.0006     &    0.02895 $\pm$ 0.00080      \\
    \hline
\end{tabular}}
\end{table}

\subsection{O-C plots}
\begin{figure}
\includegraphics [width=12cm]{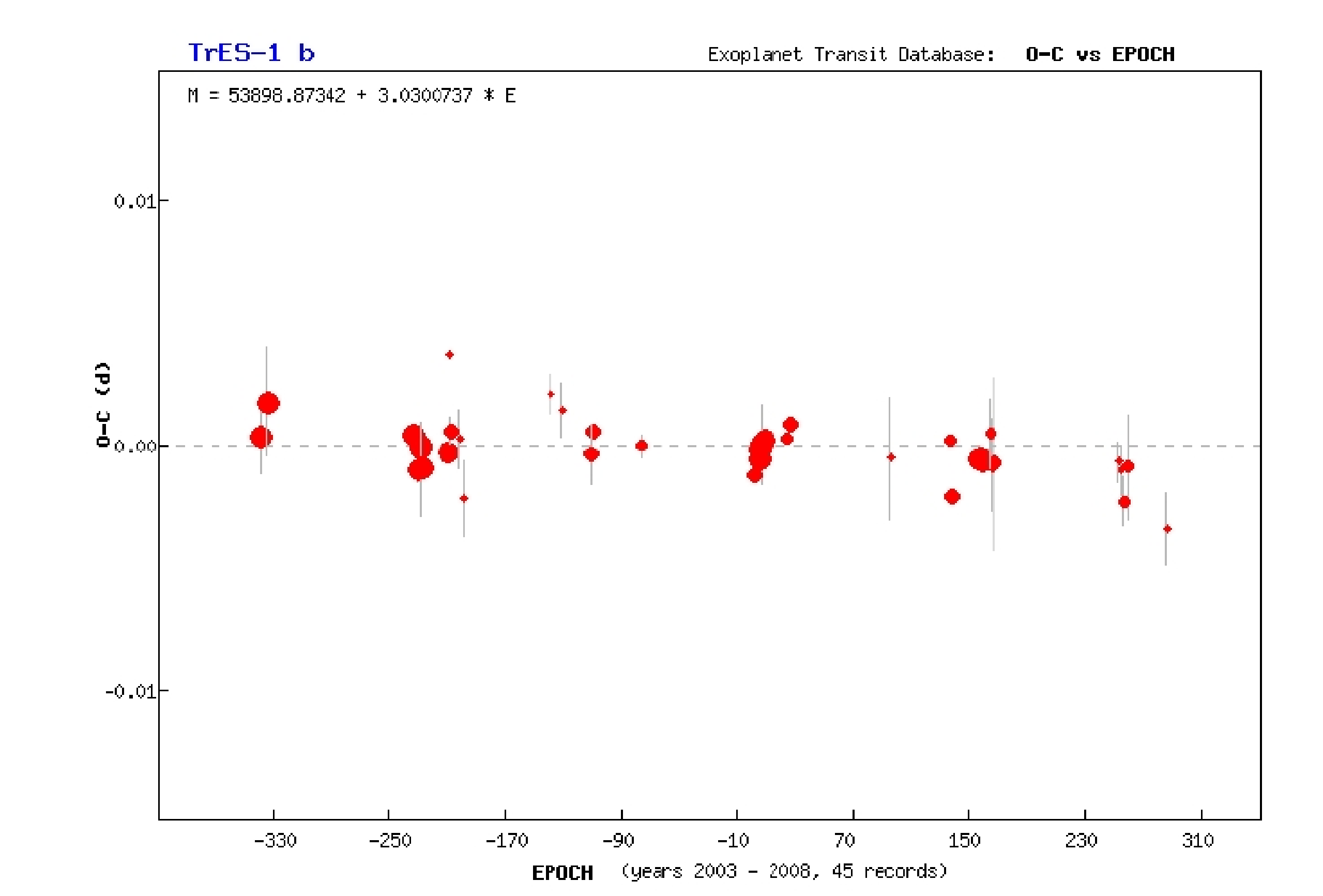} \caption{The central transit time
O-C plot for the exoplanet TrES-1 b in the ETD.\label{oc}}
\end{figure}

The O-C plots section contains the diagrams of the central transit
time, the duration of the transit and its depth as a function of
time. All available data for the selected planet including the error
bars are visible in these graphs (Fig. \ref{oc}). The quality of
individual light curve is indicated by the size of the dot. The
records in the ETD are divided into 5 groups according to their data
quality index $DQ$. While computing $DQ$ index of the light curve,
the following relation is used
\begin{equation}
   \alpha=\frac{\delta}{S} \sqrt{\rho}
      \label{rov1}
\end{equation}
where $\alpha$ is a temporary data quality index, $S$ is the mean
absolute deviation of the data from our fit and $\rho = N/l$ means
the data sampling, where $l$ is the length of observing run in
minutes.

\begin{figure}
\includegraphics [width=12cm]{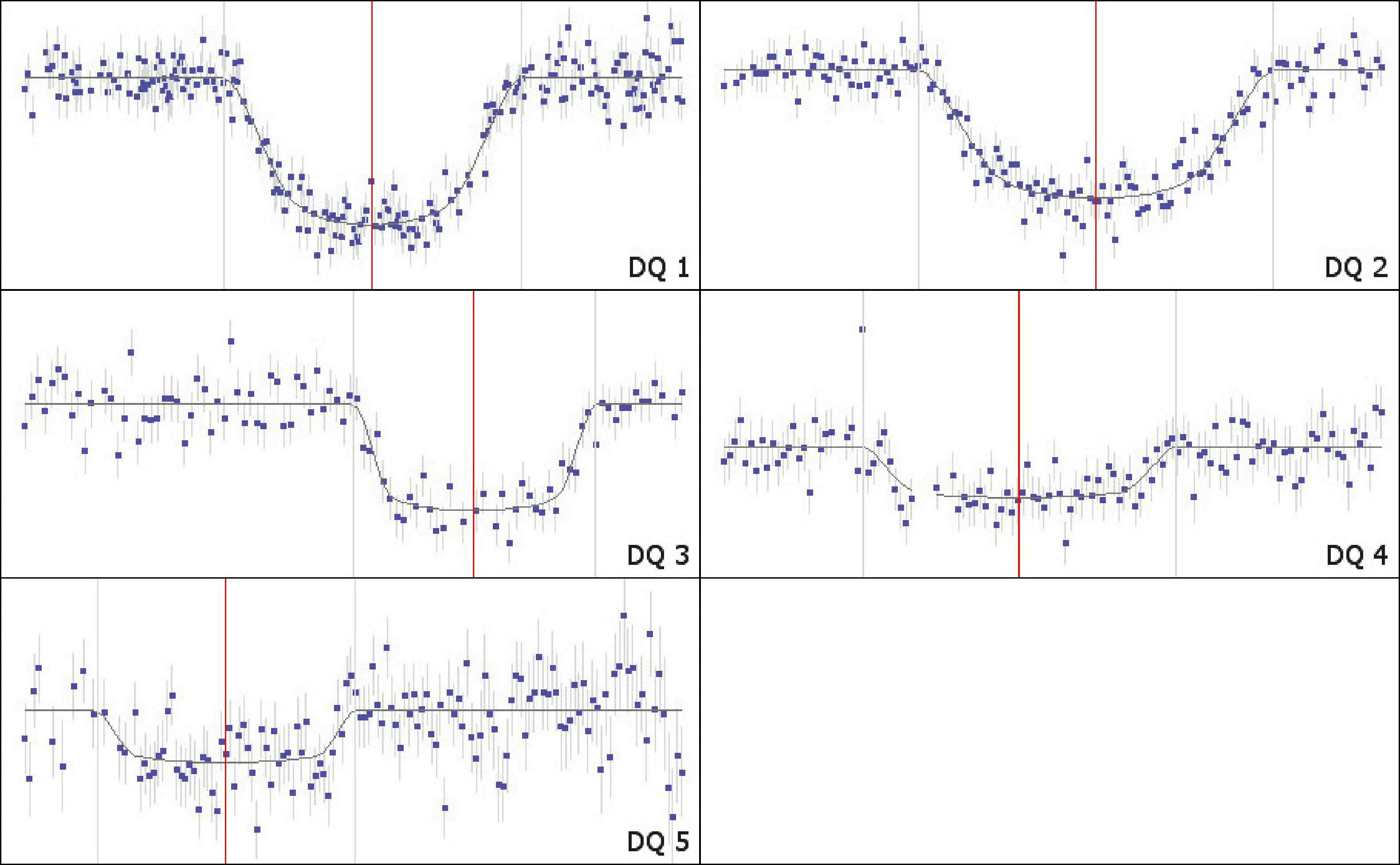} \caption{Examples of the light curve
with the different
      DQ index.\label{DQ}}
\end{figure}

The number $\alpha$ is further transformed for better lucidity to
the scale from 1 to 5 where 1 presents the best quality data and the
value 5 the worst data (Tab. \ref{Table3}, Fig. \ref{DQ}). These
thresholds are used only if whole transit is observed or when we
take over whole light curve (AXA, TRESCA). If some part (egress or
ingress) is missing in the dataset, the observation automatically
gets the worst $DQ$ index and in the notice column in the summary
table the notice "\emph{Only partial transit}" is generated. When we
take over only the results of midtransit time, transit depth and
length of the transit (not the whole light curve) we usually give
the $DQ$ equal 1.

\begin{table}
\caption{The distribution of the quality of the light curves
according to their $DQ$ index (example of the light curves see
Figure \ref{DQ}).}
 \label{Table3} \centering \scalebox{1}{
\begin{tabular}{c c c c c c}
\hline \hline \noalign{\smallskip}
    $DQ$ index & 1 & 2 & 3 & 4 & 5 \\
    \noalign{\smallskip}
    threshold & $\alpha \geq 9.5$ & $9.5 > \alpha \geq 6.0$ & $6.0 > \alpha \geq 2.5$ & $2.5 > \alpha \geq 1.3$ & $1.3 > \alpha$ \\
    \hline
\end{tabular}}
\end{table}

\subsubsection{How to download data}
In the O-C plots section of the ETD you can also download the data
for your further studies. If there is displayed whole light curve
you can download it using link in the DQ column. Whole table
including the O-C residuals can be downloaded directly using link
"\emph{Show data as ASCII table separated by semicolon}" below the
table. If the transit observation that you used in your next study
was captured to ETD from literature, you can find reference in
column "Author \& REFERENCE", so the source paper should be cited in
common way. If transit observation was published in some on-line
source (AXA, TRESCA), ask observers for a permission and other
useful comments about the data. We can supply you with e-mail
contact to observer if necessary.

\section{Future development and discussion}
We have created the on-line portal to supply observers with
information like transit predictions, TTV, variation of depth and
duration plots with availability to draw user observation to the
plot. We think that main part of the database is now ready to use
and should be useful tool for community.

In future we plan some improvements of the database to be more
user-friendly. We also plan to implement limb darkening tables into
our fits. Further in the future we would like to develop a
(semi)automatic procedure for searching of the transit timing
variations in the O-C diagrams.

\section{Acknowledgments}
The Exoplanet Transit Database is maintained by the Czech
Astronomical Society (CAS), an almost 100 years old astronomical
society with hundreds of members - professional and amateur
astronomers in the Czech Republic. ETD is stored on the web server
of CAS which is periodically backup. The server and CAS itself are
supported by grants by Czech national Council of Scientific
societies and membership fees. The long history of our organization,
large membership platform and financially assured operation of the
server are the basic conditions making the ETD permanent source
where observers can store their data securely. This investigation
was supported by the Grant Agency of the Czech Republic, grant No.
205/08/H005. We also acknowledge the support from the Research
Program MSM0021620860 of the Ministry of Education.

\end{document}